\documentclass[11pt]{article}
\usepackage{geometry}                
\usepackage{graphicx}
\usepackage{amssymb}
\usepackage{epstopdf}
\usepackage{amsmath}
\usepackage{amssymb}
\usepackage{latexsym}
\begin{document}

\begin{titlepage}
\begin{center}
{\Large\bf Tutorial on Scale and Conformal Symmetries in Diverse Dimensions}\footnote{Sodano Fest, Perugia, Italy, January 2011.}
\vskip 2ex

R. Jackiw and S.-Y. Pi\\[.5ex]
{\small\it Center for Theoretical Physics\\[.5ex]
\small\it Massachusetts Institute of Technology\\[.5ex]
\small\it Cambridge, MA 02139\\[1.5ex]
}
{\small MIT/CTP-4210}
\vskip 10ex

\begin{abstract}
We review the relation between scale and conformal symmetries in various models and dimensions. We present a dimensional reduction from relativistic to non-relativistic conformal dynamics.
\end{abstract}
\end{center}
\end{titlepage}


\section{Introduction}\label{rjSec1}
These days mathematical physicists are closely investigating scale and conformal transformations. Familiarity with theories that are invariant against these transformations --- at least on the classical, pre-quantized level --- extends for over a hundred years. Nevertheless there remain features, not unkown to some \cite{Deser1984396}, that generally have fallen into obscurity. Therefore we take this occasion of an anniversary meeting to bring into light some of these topics.

We shall describe the relation between scale and conformal symmetries and and call attention to the dimensional peculiarities and universalities of  scale and conformal transformations.

Consider a multicomponent field $\Phi$, transforming conventionally under Poincar\'{e} transformations, which leave invariant the action for $\Phi$,
\begin{eqnarray}
\mbox{translations} &:&\delta^\sigma_T \Phi (x) = \partial^\sigma \Phi(x)
\label{jackiw-eq1}\\
\mbox{Lorentz rotations} &:& \delta^{\sigma\tau}_L \Phi (x) = (x^\sigma \partial^\tau - x^\tau \partial^\sigma + \Sigma^{\sigma \tau})\ \Phi(x)
\label{jackiw-eq2}
\end{eqnarray}
where $\Sigma^{\sigma \tau}$ represents  Lorentz rotations on the components of $\Phi$; {\it i.e.} $\Sigma^{\sigma \tau}$is the spin matrix appropriate to $\Phi$. 

In addition to the Poincar\'{e} transformations we consider further transformations.
\vspace{2ex}

Scale transformations (dilations):
\begin{equation}
 \delta_S\, \Phi (x) = (x^\tau \partial_\tau + d)\ \Phi (x)
 \label{jackiw-eq3}
\end{equation}
$d$ is the scale dimension of the field $\Phi$. (We assume that the $\Phi$-multiplet carries the same scale dimension for all its components.)  $d$ is chosen so that the kinetic term in the action for $\Phi$ is invariant against dilations. We shall be concerned only with bosonic scalar and vector fields, whose kinetic term is bilinear in fields and derivatives. This is scale invariant in $D$-dimensional space-time when
\begin{equation}
d = \frac{D-2}{2} \ .
\label{jackiw-eq4}
\end{equation}
This scale dimension correctly matches the power-law behavior of the  correlation function for a free massless field: $<\Phi (x) \Phi (y)> \sim \int d^D \, k\, e^{i k (x-y)} \, /k^2 \sim (x-y)^{-D +2}$.

The additional transformations that we examine comprise the special conformal, acting on $\Phi$ as
\begin{alignat}{2}
\text{special conformal}: \quad & \delta^\sigma_C \, \Phi (x) = (2 x^\sigma \, x^\tau - g^{\sigma \tau} x^2)\, \partial_\tau \, \Phi (x) \nonumber\\[.5ex]
             			       & +2 x_\tau \, (g^{\tau \sigma} \, d - \Sigma^{\tau \sigma})\, \Phi (x)
\label{jackiw-eq5}
\end{alignat}
$g^{\alpha \beta} = \text{diag}\ (1, - 1, \ldots)$.
Fields that respond to conformal transformations by \eqref{jackiw-eq5} are called ``primary." There also exist non-primary fields,  for example the derivative of a primary field is non-primary.


 Poincar\'{e} transformations are described by the ISO(D$-$1,1) group, with $D(D+1)/2$ generators. With the addition of scale and $D$ conformal transformations, the group becomes $S O (D,2)$ with $(D+1)(D+2)/2$ generators. 
 
 The first topic that we address in Sections \ref{rjSec1.1} and \ref{rjSec1.2} concerns the relationship between dilations and conformal transformations. Specifically when dynamics is invariant against these transformations, we consider the following questions:  What are the conserved currents? What is their relation to the energy-momentum tensor $\theta^{\mu \nu}$, which is conserved and symmetric owing to Poincar\'{e} invariance?
 
 Section \ref{rjSec2} is devoted to our second topic: a dimensional reduction of the conformal group, which results in the non-relativistic conformal group.
 
 \subsection{\underline{Relation between scale and conformal symmetries}}\label{rjSec1.1}
Group structure requires that a  Poincar\'{e} invariant theory that is also conformally invariant necessarily enjoys scale invariance. This is seen from the Lie algebra, which follows from \eqref{jackiw-eq1}-\eqref{jackiw-eq3},  \eqref{jackiw-eq5}.
 \begin{equation}
[\delta^\sigma_T, \delta^\tau_C] = -2\, g^{\sigma\tau}\, \delta_S + 2\, \delta^{\sigma \tau}_L
\label{jackiw-eq6}
\end{equation}
But group theory does not establish conformal invariance from scale and Poincar\'{e} invariance; rather one must look at dynamics, which for our purpose is described by a Lagrange density depending on $\partial_\mu \Phi \ \text{and}\ \Phi: \mathcal{L} (\partial_\mu \Phi, \Phi)$.

It was shown some time ago \cite{Coleman1971552}, \cite{treiman1985} that conformal invariance holds in a Poincar\'{e} and scale invariant theory provided the quantity
\begin{subequations}\label{jackiw-eq7}
\begin{equation}
V^\alpha = \frac{\partial \mathcal{L}}{\partial \partial^\mu \Phi} \ (g^{\mu \alpha}\, d - \Sigma^{\mu \alpha}) \, \Phi
\label{jackiw-eq7a}
\end{equation}
is a total divergence.
\begin{equation}
V^\alpha = \partial_\mu \, \sigma^{\mu \alpha}.
\label{jackiw-eq7b}
\end{equation}
\end{subequations}
We call $V^\alpha$ the ``field virial."  [A related structure is seen in \eqref{jackiw-eq5}.] Moreover when \eqref{jackiw-eq7b} holds it is always possible to improve a symmetric and conserved energy momentum tensor so that it becomes traceless \cite{Coleman1971552}, \cite{treiman1985}.
\begin{equation}
\theta^{\mu \nu} \to \theta^{\mu\nu}_{CCJ}, \ \  \ g_{\mu\nu}\,  \theta^{\mu\nu}_{CCJ} = 0
\label{jackiw-eq8}
\end{equation}
where $\theta^{\mu\nu}_{CCJ} $ is defined in \cite{Callan197042}.

With this improved traceless tensor, the conserved scale and conformal currents are constructed as follows. Upon naming the infinitesimal coordinate change as $ -f^\mu (x)$ 
\begin{eqnarray}
\delta x^\mu =  -f^\mu (x)\hspace{1.5in} \nonumber\\[.5ex]
f^\mu (x) =  a^\mu,\, \omega^{\mu \alpha}\, x_\alpha \ \ (\omega^{\mu\alpha} = - \omega^{\alpha\mu}), \, c x^\mu,\, 2 c_\alpha\,  x^\alpha\, x^\mu - c^\mu\, x^2
\label{jackiw-eq9}
\end{eqnarray}
for translations, Lorentz rotations, scale and conformal transformations, respectively, the currents take the Bessel-Hagen form.
\begin{equation}
J^\mu_f = \theta^{\mu\nu}_{CCJ}\, f_\nu
\label{jackiw-eq10}
\end{equation}
To verify conservation, we use the fact that the $f^\mu$ in \eqref{jackiw-eq9} are conformal Killing vectors,
\begin{equation}
\partial_\mu f_\nu + \partial_\nu f_\mu = \frac{2}{D} \ g_{\mu \nu}\, \partial_\alpha\, f^\alpha
\label{jackiw-eq11}
\end{equation}
and $\theta^{\mu\nu}_{CCJ}$ is conserved, symmetric and traceless. 

However, if a theory is scale invariant but not conformal invariant, {\it viz.} the field virial does not satisfy \eqref{jackiw-eq7b}, then no traceless energy-momentum tensor can be constructed and the conserved scale current cannot be presented solely with the energy-momentum tensor.

To summarize: in a Poincar\'{e} invariant theory conformal transformations are symmetries provided two conditions are met.
\begin{itemize}
\item[(i)] 
Scale symmetry must hold --- this is a group theoretic requirement

\item[(ii)] 
The field virial $V^\alpha$ must be a total divergence as in \eqref{jackiw-eq7} --- this is a dynamical requirement.
\end{itemize}

It happens that in many scale non-invariant models the field virial does satisfy \eqref{jackiw-eq7b}. So that the only obstacle to conformal invariance is the absence of scale invariance. This has occasionally led to the incorrect  inference that scale invariance implies conformal invariance. This is not generically true for reasons given above. To make this point forcefully we now construct various scale invariant models that are not conformally invariant because the field virial does not satisfy \eqref{jackiw-eq7b}.

\subsection{\underline{Scale invariant but conformally non-invariant models}}\label{rjSec1.2}
\subsubsection*{A. Scalar field}
In $D$ dimensions  a single scalar field $\varphi$, governed by a Lagrangian  with the structure 
\begin{equation}
 \mathcal{L} (\partial_\mu\, \varphi, \, \varphi) = {\mathrm{L}}\, \left(\frac{\partial_\mu\, \varphi\, \partial^\mu\, \varphi}{\varphi^{\frac{2D}{D-2}}}\right) \varphi^{\frac{2D}{D-2}}
\label{jackiw-eq12}
\end{equation}
is scale invariant for arbitrary function ${\mathrm{L}}(z)$, but is conformally invariant only when ${\mathrm{L}}(z)$ is linear and independent of $z$, corresponding to a conventional  kinetic term and self interaction. For other forms of ${\mathrm{L}}$ the field virial does not satisfy \eqref{jackiw-eq7b} and conformal invariance is absent. However forms for ${\mathrm{L}}(z)$ other than ${\mathrm{L}}_0 + {\mathrm{L}}_1 z$ lead to dynamics with unconventional kinetic terms and doubts can be expressed about consistency of such unconventional dynamics, especially within quantum mechanics \cite{Polchinski1988226}. Therefore we now present a completely conventional and physically transparent model that shows similar behavior as regards to scale and conformal symmetry.

\subsubsection*{B.\ Maxwell Field}
Consider free Maxwell theory in $D$-dimensional space-time, (describing upon quantization $D - 2$ non-interacting photons). The theory is formulated in terms of a field strength $F_{\alpha\beta} = -F_{\beta\alpha}$, satisfying the equations
\begin{alignat}{2}
\text{motion:} \qquad\ & \partial_\alpha\, F^{\alpha\beta} = 0 \label{jackiw-eq13}\\[1ex]
\text{Bianchi:} \qquad\ &\partial_\alpha\, F_{\beta\gamma} + \partial_\beta\, F_{\gamma\alpha} + \partial_\gamma\, F_{\alpha\beta} = 0\label{jackiw-eq14}
\end{alignat}
These imply that the symmetric, conserved energy-momentum tensor takes the Maxwell form in any dimension,
\begin{eqnarray}
\theta^{\mu\nu} &=& - F^{\mu  \alpha} \, F^\nu_{\ \alpha} + \frac{g^{\mu\nu}}{4}\ F^{\alpha \beta}\, F_{\alpha \beta} \nonumber\\
\partial_\mu \, \theta^{\mu\nu} &=& 0
\label{jackiw-eq15}
\end{eqnarray}
but is traceless only in four dimensions.
\begin{equation}
\theta^\mu_{\ \mu} = (-1 + D/4) \ F^{\alpha \beta} \, F_{\alpha \beta}
\label{jackiw-eq16}
\end{equation}
There does not appear any way to ``improve" this into a traceless expression (as is possible in the scalar field case; but see Section \ref{rjSec1.2}\! C), so we anticipate interesting behavior with scale and conformal transformations.

The conventional approach to symmetries via Noether's theorem, {\it etc.} requires an action/Lagrangian formulation. This is achieved for the Maxwell theory by solving the Bianchi condition \eqref{jackiw-eq14} in terms of a vector potential $A_\alpha$.
\begin{equation}
F_{\alpha\beta} = \partial_\alpha\, A_\beta - \partial_\beta \, A_\alpha
\label{jackiw-eq17}
\end{equation}
$A_\alpha$ is taken to be the fundamental dynamical varialble. Then the equation of motion \eqref{jackiw-eq13} follows by varying the Lagrange density
\begin{equation}
\mathcal{L} = -\frac{1}{4}\ F^{\alpha  \beta}\, F_{\alpha \beta} = -\frac{1}{2}\ \partial^\alpha\, A^\beta\ (\partial_\alpha\, A_\beta -\partial_\beta\, A_\alpha)
\label{jackiw-eq18}
\end{equation}
with respect to $A_\alpha$. Thus the scale and conformal transformations act on $A_\alpha$. (For a different approach in $D=3$, see Section \ref{rjSec1.2}\! C.)

For dilations we adopt the following transformation laws, according to \eqref{jackiw-eq3} and \eqref{jackiw-eq4},
\begin{alignat}{2} 
&\delta_S\, A_\alpha (x) = \left(x^\tau \, \partial_\tau + \frac{D-2}{2}\right)\ A_\alpha (x) \label{jackiw-eq19}\\[1ex]
&\delta_S F_{\alpha \beta} (x) = \left(x^\tau \, \partial_\tau + \frac{D}{2}\right) F_{\alpha \beta} (x) 
 \label{jackiw-eq20}
\end{alignat}
which leave the action invariant.
\begin{equation}
\delta \, \mathcal{L} = \partial_\mu \, (x^\mu \, \mathcal{L})
\label{jackiw-eq21}
\end{equation}
Thus dilations are symmetries of the free Maxwell theory.  A conserved dilation current is constructed by Nother's theorem and an ``improvement." 
\begin{eqnarray} %
j^\mu_S &=& \frac{\partial \mathcal{L}}{\partial \partial_\mu A_\alpha}\ \delta_S\, A_\alpha - x^\mu \mathcal{L}\nonumber\\
                &=& -F^{\mu \alpha} \, (x^\beta\, F_{\beta \alpha} + x^\beta \, \partial_\alpha\, A_\beta + \frac{D-2}{2}\, A_\alpha) - x^\mu\, \mathcal{L}\nonumber\\
                &=& \theta^\mu_{\ \alpha}\, x^\alpha + \frac{4-D}{2} \, F^{\mu \alpha}\, A_\alpha + \partial_\alpha (F^{\alpha\mu}\, x^\beta\, A_\beta)
                \label{jackiw-eq22}
\end{eqnarray}
To arrive at the last line, we use the definition \eqref{jackiw-eq15} for $\theta^{\mu\nu}$ and the equation of motion \eqref{jackiw-eq13}.
We recognize that the current \eqref{jackiw-eq22} may be improved by dropping the last term in \eqref{jackiw-eq22}, because it is a divergence of an anti-symmetric tensor, hence trivially conserved. Thus we arrive at the final form for the scale current in $D$-dimensional free Maxwell theory.
\begin{equation}
J^\mu_S (x) = \theta^\mu_{\ \alpha} (x)\, x^\alpha + \frac{4-D}{2}\, F^{\mu\alpha} (x) \, A_\alpha (x)
\label{jackiw-eq23}
\end{equation}
That $J^\mu_S (x)$ is conserved follows from \eqref{jackiw-eq15} and  \eqref{jackiw-eq16}. 

Several noteworthy features characterize the scale current. First, $J^\mu_S $ is gauge variant.
\begin{eqnarray}
A_\alpha \to A_\alpha + \partial_\alpha\, \Omega &\Rightarrow& J^\mu_S \to J^\mu_S + \frac{4-D}{2} \ F^{\mu \alpha}\, \partial_\alpha\, \Omega\nonumber\\[.5ex]
                								    & =& J^\mu_S + \partial_\alpha \, \left(\frac{4-D}{2}\ F^{\mu \alpha} \Omega\, \right)
\label{jackiw-eq24}
\end{eqnarray}
But the gauge change introduces a trivially conserved term, which will not affect the dilation charge (provided surface terms are irrelevant).
\begin{equation}
Q_S \to Q_S + \int \, d^{D-1} x \, \partial_i \ \left(\frac{4-D}{2}\ F^{0 i}\, \Omega\right) = Q_S
\label{jackiw-eq25}
\end{equation}
Second, the $D\ne 4$ scale current involves terms additional to the energy momentum tensor, which are needed when $\theta^\mu_{ \ \mu}$ fails to vanish. Indeed as will be seen presently, the last term in $J^\mu_S$ is precisely the Maxwell theory field virial. That $\theta^\mu_{\ \mu}$ is non-vanishing and that the field virial is not a total divergence indicates that conformal symmetry is absent in the above, vector-potential formulated, Maxwell theory. 

The conformal transformation \eqref{jackiw-eq5} when specialized to a vector potential, as a primary field $\Phi \to A_\mu$, reads
\begin{multline}
\delta^\sigma_C\, A_\alpha (x) = (2 x^\sigma \, x^\tau - g^{\sigma \tau}\, x^2)\ \partial_\tau \, A_\alpha (x) + (D-2)\, x^\sigma\, A_\alpha (x)\\[.5ex]
             - 2\, x_\alpha \, A^\sigma (x) + 2\, g^\sigma_\alpha\, x^\tau\, A_\tau (x). \hspace{2.55in} 
             \label{jackiw-eq26}
\end{multline}
For the field strength $F_{\alpha\beta}$ \eqref{jackiw-eq26} implies
\begin{eqnarray}
\delta^\sigma_C\, F_{\alpha\beta} (x) &=& (2 x^\sigma \, x^\tau - g^{\sigma \tau}\, x^2) \ \partial_\tau F_{\alpha\beta} (x)\nonumber\\[.5ex]
                                                                    && + D \, x^\sigma F_{\alpha\beta} (x) + 2\, g^\sigma_\alpha \, x^\tau \, F_{\tau \beta} (x) + 2\, g^\sigma_\beta\, x^\tau\, F_{\alpha \tau} (x)\nonumber\\[.5ex]
                                                                     && -2 x_\alpha\, F^\sigma_{\ \beta} (x) - 2 x_\beta\, F^{\ \sigma}_\alpha (x) + (D-4) (g^\sigma_\alpha\, A_\beta (x) - g^\sigma_\beta\, A_\alpha (x )
\label{jackiw-eq27} 
\end{eqnarray}

Because $F_{\alpha\beta}$ involves derivatives of the primary field $A_\alpha$, the transformation rule \eqref{jackiw-eq27} describes a non-primary field for $D \ne 4$. If $F_{\alpha\beta}$ were primary it would transform, according to \eqref{jackiw-eq5} with $\Phi \to F_{\alpha\beta}$, as
\begin{eqnarray}
\triangle^\sigma_C\, F_{\alpha \beta} & =& (2 x^\sigma \, x^\tau - g^{\sigma \tau}\, x^2)\ \partial_\tau\, F_{\alpha \beta} + D\, x^\sigma\, F_{\alpha \beta}\nonumber\\[.5ex]
 && + 2\, g^\sigma_\alpha \, x^\tau \, F_{\tau \beta} + 2\, g^\sigma_\beta\, x^\tau \, F_{\alpha \tau} - 2\, x_\alpha\, F^{\sigma}_{\  \beta} - 2\, x_\beta\, F^{\  \sigma}_{\alpha}
 \label{jackiw-eq28}
\end{eqnarray}
Evidently
\begin{equation}
\delta^\sigma_C\, F_{\alpha\beta} = \triangle^\sigma_C\, F_{\alpha \beta} + (D-4) (g^\sigma_\alpha\, A_\beta - g^\sigma_\beta\, A_\alpha)
\label{jackiw-eq29}
\end{equation}

The transformation rule \eqref{jackiw-eq27} implies that the Lagrange density \eqref{jackiw-eq18} changes as
\begin{equation}
\delta^\sigma_C\, \mathcal{L} = \partial_\mu \ \left[(2 x^\sigma\, x^\mu - g^{\sigma \mu}\, x^2) \mathcal{L})\right] + (4-D)\ F^{\sigma \tau}\, A_\tau
\label{jackiw-eq30}
\end{equation}
The last contribution shows that the action is not invariant at $D\ne4$. The same conclusion is reached from the field virial, which here reads
\begin{equation}
V^\alpha = \frac{4-D}{2}\ F^{\alpha \beta}\, A_\beta
\label{jackiw-eq31}
\end{equation}
--- an expression, which we have seen previously forming part of the scale current \eqref{jackiw-eq23}. Since $V^\alpha \ne \partial_\beta\, \sigma^{\alpha \beta}$, conformal symmetry is absent for $D\ne 4$.

One may ask whether the conformal transformation leaves invariant the equations of motion, even though the action is not invariant. For this to be true $\partial_\alpha\, \delta^\sigma_C \, F^{\alpha\beta}$ must vanish. But it does not; from \eqref{jackiw-eq25} we have
\begin{equation}
\partial_\alpha \, \delta^\sigma_c\, F^{\alpha \beta} = (D-4) (\partial^\beta\, A^\sigma - g^{\beta\sigma}\, \partial_\tau \, A^\tau) \ne 0.
\label{jackiw-eq32}
\end{equation}
\noindent
[A curious observation is that if we adopt for $F_{\alpha\beta}$ (without justification) the primary field transformation rule $\triangle^\sigma_C\, F_{\alpha\beta}$ in \eqref{jackiw-eq28}, then 
\begin{equation}
\triangle^\sigma_C \, \mathcal{L}  = \partial_\tau\, \left[\left(2 x^\sigma \, x^\tau - g^{\sigma \tau}\, x^2\right) \ \mathcal{L}\right] 
\label{jackiw-eq33}
\end{equation}
and the action is invariant. But this does not lead to a conserved current, because $\triangle^\sigma_C\, F_{\alpha\beta}$ cannot be obtained from a transformation on a vector potential, the fundamental variable, whose variation leads to the equation of motion \eqref{jackiw-eq13}. In other words, $\triangle^\sigma_C\, F_{\alpha\beta}$ does not respect the equation of motion nor the Bianchi identity. ]

Note that the Poincar\'{e} and dilation currents can be presented as
\begin{equation}
J^\mu_f = \theta^{\mu\alpha}\, f_\alpha + \frac{4-D}{2D}\ \partial_\alpha\, f^\alpha\, F^{\mu \beta}\, A_\beta
\label{jackiw-eq34}
\end{equation}
where $f^\mu$ is the coordinate transformation, defined in \eqref{jackiw-eq9}. The divergence of \eqref{jackiw-eq34} reads
\begin{eqnarray}
\partial_\mu \, J^\mu_f &=& \theta^\mu_{\ \mu}\, \frac{\partial_\alpha f^\alpha}{D} + \frac{4-D}{4D}\ \partial_\alpha \, f^\alpha\, F^{\mu\beta}\, F_{\mu\beta}\nonumber\\[.5ex]
                                          && +\frac{4-D}{2D}\ \partial_\mu\, \partial_\alpha \, f^\alpha\, F^{\mu\beta}\, A_\beta
                                          \label{jackiw-eq35}
\end{eqnarray}
where the Killing equation \eqref{jackiw-eq11} and the field equation \eqref{jackiw-eq13} are used. The first two terms on the right in \eqref{jackiw-eq35} cancel; in the last term $\partial_\mu\, \partial_\alpha \, f^\alpha$ vanishes for Poincar\'{e} and scale transformations. But for conformal translations $\partial_\mu\, \partial_\alpha \, f^\alpha = 2 D c_\mu $ and $\partial_\mu\, J^\mu_C = (4-D) c_\mu\, F^{\mu\beta}\, A_\beta$ in agreement with \eqref{jackiw-eq30}. Thus we may consider \eqref{jackiw-eq34} as a generalization of the Bessel-Hagen formula, producing conserved Poincar\'{e} and scale currents and a non-conserved conformal current when $D \ne 4$..

\vspace{6ex}

One further observation on the response of a vector field to the transformations \eqref{jackiw-eq9} is that the infinitesimal change of $A_\alpha$ does not coincide with the Lie derivative when $D\ne 4$. 
Rather we have
\begin{eqnarray}
& \delta_f \, A_\alpha = \mathrm{L}_f\, A_\alpha + \frac{D-4}{2D}\ \partial_\mu\, f^\mu\, A_\alpha
\label{jackiw-eq36}
\end{eqnarray}
\vspace{-1.5ex}
where
\vspace{-1.5ex}
\begin{eqnarray}
& \mathrm{L}_f\, A_\alpha = f^\mu\, \partial_\mu\, A_\alpha + \partial_\alpha\, f^\mu\, A_\mu
\label{jackiw-eq37}
\end{eqnarray}
$\partial_\mu f^\mu$ vanishes for Poincar\'{e} transformations, but survives for scale and conformal transformations, and moves the transformation law away from the Lie derivative, thereby assigning to $A_\mu$ the scale dimension $\frac{D-2}{2}$. In contrast, geometric arguments would assign unit scale dimension to a vector potential, which naturally belongs with the unit-dimensional derivative. Also when $\partial_\mu\, f^\mu \ne 0 \ \text{and}\ D\ne 4$ the transformation law \eqref{jackiw-eq36} interferes with gauge covariance: one expects $\delta_f A_\mu$ to be gauge invariant up to a gauge transformation. The Lie derivative \eqref{jackiw-eq37} preserves this requirement, because it may be presented as
\begin{equation}
\mathrm{L}_f A_\alpha = f^\mu\, F_{\mu\alpha} + \partial_\alpha \ (f^\mu\, A_\mu)
\label{jackiw-eq38}
\end{equation}
But the term in $\delta_f \, A_\mu$ beyond $\mathrm{L}_f A_\alpha$ is gauge non-invariant.

\subsubsection*{C. Secret Conformal Symmetry of $D=3$ Maxwell Theory}
We have seen that conformal symmetry is incompatible with a vector potential formulation of Maxwell theory in  $D\ne 4$. But in $D=3$, a scalar potential formulation is available and conformal invariance can be implemented. To achieve an action/Lagrangian formulation of the free Maxwell theory, we may present in $D=3$ the anti-symmetric $F_{\alpha\beta}$  as 
\begin{subequations}\label{jackiw-eq39}
\begin{eqnarray}
F_{\alpha\beta} = \varepsilon_{\alpha\beta\mu} \, \partial^\mu\, \varphi \label{jackiw-eq39a}\\[.5ex]
\partial^\mu\, \varphi = \frac{1}{2}\, \varepsilon^{\mu\alpha\beta}\, F_{\alpha\beta}\label{jackiw-eq39b}
\end{eqnarray}
\end{subequations}
Now $\varphi$  is the fundamental dynamical variable, and it describes the single ``photon" degree of freedom in $D=3$, without the redundancy of a vector potential.

The Maxwell field equation \eqref{jackiw-eq13} becomes the identity $\varepsilon_{\alpha\beta\mu} \, \partial^\beta\, \partial^\alpha\, \varphi = 0$, while the Maxwell Bianchi identity becomes the free field equation for $\varphi$. 
\begin{equation}
\square\, \varphi = \frac{1}{2}\ \partial_\mu\, \varepsilon^{\mu\alpha\beta}\, F_{\alpha\beta} = 0
\label{jackiw-eq40}
\end{equation}
Such an interchange is characteristic of a dual relation. (In a Euclidean formulation this is analogous to presenting a source free, three-dimensional magnetic field in terms of a magnetic scalar potential.)

The Lagrange density now reads
\begin{equation}
\mathcal{L} = -\frac{1}{4}\ F^{\alpha \beta}\, F_{\alpha \beta} = -\frac{1}{2}\ \partial^\mu\, \varphi\, \partial_\mu\, \varphi
\label{jackiw-eq41}
\end{equation}
and the dynamics enjoyed by $\varphi$ leads to the conventional, conserved and symmetric energy-momentum tensor
\begin{equation}
\theta^{\mu\nu} = \partial^\mu\, \varphi\, \partial^\nu\, \varphi - \frac{1}{2}\  g^{\mu\nu} \partial^\alpha \, \varphi\, \partial_\alpha\, \varphi
\label{jackiw-eq42}
\end{equation}
which may be improved, so it is traceless.
\begin{eqnarray}
\theta^{\mu\nu}_{CCJ} &=& \theta^{\mu\nu} + \frac{1}{8}\ (g^{\mu\nu}\ \square - \partial^\mu\, \partial^\nu)\ \varphi^2\nonumber\\
g_{\mu\nu} \, \theta^{\mu\nu}_{CCJ} &=& 0
\label{jackiw-eq43}
\end{eqnarray}
The scale transformation on $\varphi$ follows \eqref{jackiw-eq3} and  \eqref{jackiw-eq4} [with $D=3$]
\begin{equation}
\delta_S\, \varphi (x) = \left(x^\tau\, \partial_\tau + \frac{1}{2}\right)\ \varphi (x)
\label{jackiw-eq44}
\end{equation}
while the conformal transformation reads from \eqref{jackiw-eq5}
\begin{equation}
\delta^\sigma_C\, \varphi (x) = (2 x^\sigma\, x^\tau - g^{\sigma \tau}\, x^2)\ \partial_\tau\, \varphi (x) + x^\sigma\, \varphi (x)
\label{jackiw-eq45}
\end{equation}
Eqn. \eqref{jackiw-eq45} now determines the conformal transformation rule for the field strength \eqref{jackiw-eq39a}
\begin{eqnarray}
\bar{\delta}^\sigma_C \, F_{\alpha\beta} (x) =  (2 x^\sigma\, x^\tau - g^{\sigma \tau}\, x^2) \ \partial_\tau\, F_{\alpha\beta} (x) + 3 x^\sigma\, F_{\alpha\beta} (x)\nonumber\\[.5ex]
+ 2\, g^\sigma_\alpha\, x^\tau\, F_{\tau\beta} (x) + 2 \, g^\sigma_\beta\, x^\tau\, F_{\alpha\tau} (x) -2\, x_\alpha\, F^\sigma_{\ \beta} (x) - 2 \, x_\beta\, F^{\ \sigma}_\alpha (x)
 + \varepsilon^{\ \ \ \sigma}_{\alpha \beta} \varphi (x) 
 \label{jackiw-eq46}
\end{eqnarray}
The over-bar on $\bar{\delta}^\sigma_C$ distinguishes the present, scalar potential based, transformation on $F_{\alpha\beta}$, from the earlier, vector potential based formula \eqref{jackiw-eq27}. According to the new transformation law \eqref{jackiw-eq46}, $F_{\alpha\beta}$ remains non-primary, since $\varphi$ enters the transformation. The present analog to \eqref{jackiw-eq29} is 
\begin{equation}
\bar{\delta}^\sigma_C  \, F_{\alpha\beta} = \triangle^\sigma\, F_{\alpha\beta} + \varepsilon_{\alpha_\beta}^{\ \ \ \sigma} \, \varphi
\label{jackiw-eq47}
\end{equation}
The action is invariant.
\begin{equation}
\bar{\delta}^\sigma_C\, \mathcal{L} = \partial_\tau \left[(2 x^\sigma\, x^\tau - g^{\sigma \tau}\, x^2)\ \mathcal{L} - g^{\sigma \tau}\,  \frac{\varphi^2}{2} \right]
\label{jackiw-eq48}
\end{equation}

The conformal symmetry current is constructed in the usual way from the improved, traceless energy momentum tensor \eqref{jackiw-eq43} \cite{Callan197042}, which can be presented by virtue of \eqref{jackiw-eq39} in terms of $F_{\alpha\beta}$ as
\begin{equation}
\theta^{\mu\nu}_{\ CCJ} = -\frac{3}{4}\ F^{\mu\alpha}\, F^\nu_{\ \alpha} + \frac{1}{4}\ g^{\mu\nu}\, F^{\alpha_\beta}\, F_{\alpha_\beta}\ - \frac{\varphi}{16}\ \left(\partial^\mu\, \varepsilon^{\nu \alpha\beta}\, F_{\alpha\beta} + \partial^\nu\, \varepsilon^{\mu \alpha\beta}\, F_{\alpha_\beta}\right)
\label{jackiw-eq49}
\end{equation}
Note that dependence on $\varphi$ remains.

Comparison of the vector potential based formulation with the scalar based one indicates that there is a non-local relation between $A_\alpha$ and $\varphi$, which implements the duality
\begin{equation}
\varepsilon^{\mu\alpha\beta}\, \partial_\alpha\, A_\beta \raisebox{1.5ex}{$\underset{=}{?}$}\,  \partial^\mu \varphi
\label{jackiw-eq50}
\end{equation}
Whether conformally symmetric  dual formulations for the free Maxwell theory exist in other dimensions, remains an open question. So a conformally invariant potential formulation for Maxwell theory at $D>4$ remains unknown.
\newpage

\section{Dimensional Reduction of the Conformal Group}\label{rjSec2}
The second topic that we address concerns the universality across dimensions of scale/conformal transformations/invariances. We consider an $N$ component multiplet of scalar fields $ \Phi$ with interactions that preserve scale and conformal symmetries [see \eqref{jackiw-eq12}]
\begin{alignat}{2}
&\mathcal{L} = \frac{1}{2}\ \partial_\mu\, \Phi \cdot \partial^\mu\,  \Phi - \lambda\, ( \Phi \cdot  \Phi)^{\frac{D}{D-2}}\label{jackiw-eq51}\\[.5ex] 
&\delta_S\,  \Phi (x) = x^\tau\, \partial_\tau\,  \Phi (x) + \frac{D-2}{2}\  \Phi (x) \label{jackiw-eq52} \\[.5ex] 
&\delta^\sigma_C\,  \Phi (x) = \left(2 x^\sigma \, x^\tau - g^{\sigma\tau}\, x^2\right)\ \partial_\tau\,   \Phi (x)  + (D-2) \, x^\sigma \, \Phi (\mathbf{x})\hspace{.95in} \label{jackiw-eq53}
\end{alignat}
As stated earlier, the ISO ($D-1,1$) Poincar\'{e} invariance is enlarged to the $SO(D,2)$ conformal group, and the formulas \eqref{jackiw-eq51}--\eqref{jackiw-eq53} hold for arbitrary $D>2$. The independent variable $x^\mu$ describes time at $\mu = 0$ and there are $D-1$ spatial coordinates  $\mathbf{x}$.  $D = 2$ does not fit above scheme \cite{Jackiw:2005su}, but let us continue to $D=1$. There the spatial coordinates disappear and the ``field" $ \Phi$ becomes an N-component variable --- we shall rename it $\mathbf{q}$ --- which depends on the surviving variable $t$. The Lagrange density \eqref{jackiw-eq51} at $D=1$ reduces to
\begin{equation}
L = \frac{1}{2}\ \frac{d}{dt}\ \mathbf{q} \cdot \frac{d}{dt}\ \mathbf{q} - \lambda/q^2.
\label{jackiw-eq54}
\end{equation}
The Poincar\'{e} transformations contract to time translations generated by the Hamiltonian $H$ --- the only Poincar\'{e} generator that survives in the $D=1$ limit. The reduced scale transformation \eqref{jackiw-eq52} reads
\begin{equation}
\delta_S\, \mathbf{q} (t) = t\, \partial_t\, \mathbf{q}\, (t) - \frac{1}{2}\ \mathbf{q} (t).
\label{jackiw-eq55}
\end{equation}
Only a single conformal transformation remains from \eqref{jackiw-eq53}
\begin{equation}
\delta_C\, \mathbf{q} (t) = t^2\, \partial_t\, \mathbf{q}\, (t) - t\, \mathbf{q}\, (t).
\label{jackiw-eq56}
\end{equation}

We now recognize \eqref{jackiw-eq54}--\eqref{jackiw-eq56} as a non-relativistic energy-conserving system in $N$ spatial dimensions, enjoying the most general velocity-independent scale and conformal symmetries \cite{rjack1972}. It is amusing that such mundane mechanics can be attained as limit of a relativistic, conformally invariant quantum field theory.

\section*{Acknowledgment}
We thank S. Deser and M. Shifman for discussions about the dual formulation of the $D=3$ Maxwell theory. This work is supported by the U.S. Department of Energy (DoE) under the cooperative
research agreement DE-FG02-05ER41360.

\end{document}


\title{Addendum to  ``Tutorial on Scale and Conformal Symmetries ..."}

\author{S.-H. Ho, R. Jackiw and S.-Y. Pi}

\affiliation{Center for Theoretical Physics \\
Massachusetts Institute of Technology \\
 Cambridge, MA 02139}


\begin{abstract}

When discussing with colleagues
the recently posted ``Tutorial on Scale and Conformal Symmetries in Diverse Dimensions"\cite{Jackiw:2011vz}, we developed representations for finite conformal transformations of tensor and spinor fields, in which the Lorentz tensorial/spinorial structure of the field plays no role. Since we have not seen these expressions in the literature, we present them here in case they are unknown.

\end{abstract}

\preprint{MIT/CTP-4213}

\maketitle


\subsection{Conformal transformation on coordinates and fields}

To set notation, we review well known formulas. A finite conformal transformation acts on the space-time coordinate $x^{\mu}$ with the rule
\beq 
\label{eq1}  x^{\mu} \rightarrow x^{'\mu}= \frac{x^{\mu}+c^{\mu}x^2}{\sigma(x,c)}, \\
\label{eq2}   \sigma(x,c)=1+2 c\cdot x+ c^2 x^2.
\eeq
The Abelian composition law is readily verified
with the help of
\beq \label{eq3}
\sigma(x,c)\sigma(x',c')=\sigma(x, c'+c). 
\eeq
The transformation (\ref{eq1}), (\ref{eq2}) can also be presented as
\beq \label{eq4}
x^{'\mu}=\frac{X^{\mu}+c^{\mu}}{(X+c)^2},
\eeq
 where $X^{\mu}$ is the inversion.
\beq \label{eq5}
X^{\mu} \equiv \frac{x^{\mu}}{x^2}
\eeq
The infinitesimal version of (\ref{eq1}), (\ref{eq2}) and (\ref{eq4}) reads
\beq \label{eq6}
\delta x^{\mu} \equiv x^{'\mu}-x^{\mu}=- 2 (c \cdot x) x^{\mu} +c^{\mu}x^2\equiv - f^{\mu}(x).
\eeq
The infinitesimal transformation for a field $\Phi$ is
\beq \label{eq7}
\delta \Phi(x) \equiv \Phi'(x)-\Phi(x)=f^{\mu}(x)\partial_{\mu} \Phi(x)+2 (c \cdot x) d \Phi(x) +2 c_{\mu} x_{\nu} \Sigma^{\mu \nu} \Phi(x),
\eeq 
where d is the scale dimension and $\Sigma^{\mu\nu}$ is the spin matrix for field $\Phi$. 

Construction of the finite conformal transformation for $\Phi$ proceeds by considering separately different spins. For a scalar field $S$, $\Sigma^{\mu\nu}$ is absent and (\ref{eq7}) simplifies.
\begin{subequations}\label{eq8}
\beq
\label{eq8a}   \delta S(x) \equiv S'(x)-S(x)=f^{\mu}(x)\partial_{\mu} S(x)+2 (c \cdot x) d S(x) 
\eeq
The first term on the right is the Lie derivative of a scalar $L_f S=f^{\mu}\partial_{\mu}S$. So (\ref{eq8a}) may be presented as 
\beq 
\label{eq8b}S'(x)-L_f S(x)= S(x)+2 (c \cdot x) d S(x),
\eeq
\end{subequations}
and the finite version of (\ref{eq8b}) reads
\beq \label{eq9}
S'(x')=[\sigma(x,c)]^d S(x).
\eeq
The diffeomorphism structure is a consequence of the Lie derivative and the $\sigma^d$ factor is needed to account for the last term in (\ref{eq8}).  The composition law for successive iteration of (\ref{eq9}) follows from (\ref{eq3}).

For a vector field $V_{\alpha}$, (\ref{eq7}) specializes
to 
\begin{subequations} \label{eq10}
\beq \label{eq10a}
&&\delta V_{\alpha}(x) \equiv V'_{\alpha}(x)-V_{\alpha}(x)=f^{\mu}(x)\partial_{\mu} V_{\alpha}(x)+2(c \cdot x) d V_{\alpha}(x)+n_{\alpha}^{\ \beta}(x)V_{\beta}(x), \nonumber \\
&& n_{\alpha}^{\ \beta}(x) \equiv 2 c_{\alpha}x^{\beta}-2 c^{\beta} x_{\alpha} .
\eeq
$n_{\alpha}^{\ \beta}$ arises from the spin matrix of $V_{\alpha}$. Upon isolating
the Lie derivatives of $V_{\alpha}$: $L_f V_{\alpha}=f^{\mu}\partial_{\mu} V_{\alpha}+\partial_{\mu}f^{\alpha} V_{\alpha}$, (\ref{eq10a}) becomes
\beq \label{eq10b}
V'_{\alpha}(x)-L_f V_{\alpha}(x)=V_{\alpha}(x)+ (d-1) 2(c \cdot x)V_{\alpha}(x),
\eeq
\end{subequations}
and we recognize the finite transformation.
\beq \label{eq11}
V'_{\alpha}(x')=[\sigma(x,c)]^{d-1} V_{\beta}(x) \frac{\partial x^{\beta}}{\partial x^{'\alpha}}
\eeq


\subsection{Another formula for finite conformal transformations}
Our alternative transformation rule begins with the usual expression for a spinless field, Eq.(\ref{eq9}). For a vector field, we proceed by defining the matrix
\beq \label{eq12}
\mathcal{I}_{\alpha}^{\ \beta}(x) && \equiv g_{\alpha}^{\ \beta}-2\hat{x}_{\alpha}\hat{x}^{\beta}, \nonumber \\
 \hat{x}^{\mu}&& \equiv \frac{x^{\mu}}{\sqrt{x^2}}.
\eeq
Note that 
\begin{subequations}
\beq \label{eq13a}
\mathcal{I}_{\alpha}^{\ \beta}(x)=x^2 \frac{\partial X^{\beta}}{\partial x^{\alpha}},
\eeq
and $\mathcal{I}_{\alpha}^{\ \beta}$ is an improper Lorentz matrix.
\beq 
\label{eq13b}\mathcal{I}_{\alpha}^{\ \beta}(x) \ \mathcal{I}_{\mu}^{\ \nu}(x) \ g_{\beta \nu} =g_{\alpha \mu} \\
\label{eq13c} det \ \mathcal{I}_{\alpha}^{\ \beta}(x) = -1 
\eeq
\end{subequations}
Consider now the variation of 
\beq \label{eq14}
\widetilde{V}_{\alpha}(x) \equiv \mathcal{I}_{\alpha}^{\ \beta}(x) V_{\beta}(x).
\eeq
From (\ref{eq10a}) it follows that 
\begin{subequations} \label{eq15}
\beq 
\label{eq15a}
\delta \widetilde{V}_{\alpha}&&\equiv \mathcal{I}_{\alpha}^{\ \beta}(x) \delta V_{\beta}(x)=\mathcal{I}_{\alpha}^{\ \beta}(x) f^{\mu}(x)\partial_{\mu} V_{\beta}(x) + 2 (c \cdot x)d \widetilde{V}_{\alpha}(x) +\mathcal{I}_{\alpha}^{\ \beta}(x)n_{\beta}^{\ \gamma}(x)V_{\gamma}(x) \nonumber \\
&&= f^{\mu}(x)\partial_{\mu} \widetilde{V}_{\alpha}(x)+2 (c \cdot x)d \widetilde{V}_{\alpha}(x)-\left\{f^{\mu}(x)\partial_{\mu}\mathcal{I}_{\alpha}^{\ \gamma}(x)-\mathcal{I}_{\alpha}^{\ \beta}(x)n_{\beta}^{\ \gamma}(x)\right\} V_{\gamma}(x).
 \eeq
The surprise is that the bracketed quantity vanishes, and one is left with 
 \beq \label{eq15b}
\delta \widetilde{V}_{\alpha}(x)= f^{\mu}(x)\partial_{\mu}\widetilde{V}_{\alpha}(x)+2 (c\cdot x)d \widetilde{V}_{\alpha}(x).
\eeq
\end{subequations}
The spin components have decoupled. $\widetilde{V}_{\alpha}$ behaves under conformal transformation like a scalar field. The finite transformation then follows from (\ref{eq9})
\begin{subequations} \label{eq16}
\beq \label{eq16a}
\widetilde{V}'_{\alpha}(x')=[\sigma(x,c)]^d \widetilde{V}_{\alpha}(x),
\eeq
or equivalently
\beq \label{eq16b}
V'_{\alpha}(x')=[\sigma(x,c)]^d \mathcal{I}_{\alpha}^{\ \beta}(x')\mathcal{I}_{\beta}^{\ \gamma}(x)V_{\gamma}(x).
\eeq
\end{subequations}


It remains to show that the rule (\ref{eq16}) coincides with the conventional formula (\ref{eq11}). We need to calculate $\partial x^{\beta} / \partial x^{'\alpha}$, but we begin with the inverse.
\begin{subequations} \label{eq17}
\beq \label{eq17a}
\frac{\partial x^{'\alpha}}{\partial x^{\beta}}=\frac{\partial X^{\mu}}{\partial x^{\beta}}  \frac{\partial}{\partial X^{\mu}} \frac{(X^{\alpha}+c^{\alpha})}{(X+c)^2}
\eeq
We have used the chain rule and formula (\ref{eq4}) for $x^{'\alpha}$. From (\ref{eq13a}) it follows that
\beq \label{eq17b}
\frac{\partial x^{'\alpha}}{\partial x^{\beta}}=\frac{1}{x^2} \mathcal{I}_{\beta}^{\ \mu}(x) \frac{1}{(X+c)^2} \mathcal{I}_{\mu}^{\ \alpha}(X+c).
\eeq
Using $(X+c)^2=\sigma(x,c) / x^2$ and $\mathcal{I}_{\mu}^{\ \alpha}(X+c)=\mathcal{I}_{\mu}^{\ \alpha}(x')$, we can present (\ref{eq17a}) as
\beq \label{eq17c}
\frac{\partial x^{'\alpha}}{\partial x^{\beta}}=\frac{1}{x^2} \mathcal{I}_{\beta}^{\ \mu}(x) \frac{x^2}{\sigma(x,c)} \mathcal{I}_{\mu}^{\ \alpha}(x'),
\eeq
\end{subequations}
and it follows that
\beq \label{eq18}
\frac{\partial x^{\beta}}{\partial x^{'\alpha}}= \mathcal{I}_{\alpha}^{\ \gamma}(x') \mathcal{I}_{\gamma}^{\ \beta}(x) \sigma(x,c).
\eeq
This establishes that (\ref{eq11}) coincides with (\ref{eq16b}).


\subsection{The spinor case}
The infinitesimal conformal transform of a spinor field $\psi$ follows (\ref{eq7}) with $\Sigma^{\mu\nu}=\frac{1}{4} [\gamma^{\mu},\gamma^{\nu}]$. We define a new spinor field
\beq \label{eq19}
\widetilde{\psi}(x)=\gamma \cdot \hat{x} \psi(x), 
\eeq
and deduce from (\ref{eq7}) that
\beq \label{eq20}
\delta \widetilde{\psi}(x) && \equiv \gamma \cdot \hat{x} \delta \psi(x) = \gamma \cdot \hat{x} f^{\mu}(x) \partial_{\mu} \psi(x) + 2 (c \cdot x) d \widetilde{\psi}(x)+2 c_{\mu}x_{\nu} (\gamma \cdot \hat{x}) \Sigma^{\mu\nu}\psi(x) \nonumber \\
&&= f^{\mu}(x)\partial_{\mu} \widetilde{\psi}(x)+2 (x \cdot c)d\widetilde{\psi}(x)- \{ f^{\mu}(x)\partial_{\mu} \gamma \cdot \hat{x}-2 c_{\mu}x_{\nu} (\gamma \cdot \hat{x}) \Sigma^{\mu\nu}\}\psi(x).
\eeq
As in the vector case, the bracketed term vanishes, and the spinor components decouple, leaving
\beq \label{eq21}
\widetilde{\psi}'(x')-f^{\mu}(x)\partial_{\mu}\widetilde{\psi}(x)=\widetilde{\psi}(x)+ 2(c \cdot x) d \widetilde{\psi}(x),
\eeq
or in the finite form
\beq \label{eq22}
\widetilde{\psi}'(x')=[\sigma(x,c)]^d \widetilde{\psi}(x).
\eeq
This leads to
\beq \label{eq23}
\psi'(x')&&=[\sigma(x,c)]^d (\gamma \cdot \hat{x}') (\gamma \cdot \hat{x}) \psi(x) \nonumber \\
&&=[\sigma(x,c)]^{d-1/2}(1+c_{\mu}x_{\nu}\gamma^{\mu}\gamma^{\nu})  \psi(x),
\eeq
which is the conventional expression for a finite conformal transformation of a spinor field.


\subsection{Conclusion}

For tensor fields and spinor fields  multiplication by $\mathcal{I}_{\mu}^{\ \nu}(x)$ and $\gamma \cdot \hat{x}$, respectively, decouples the Lorentz indices from the conformal transformation. The relation between the two decoupling matrices is exhibited by the formula
\beq \label{eq24}
(\gamma \cdot \hat{x}) \gamma_{\mu} (\gamma \cdot \hat{x})=-\mathcal{I}_{\mu}^{\ \nu}(x) \gamma_{\nu}.
\eeq
The matrix $\mathcal{I}_{\mu}^{\ \nu}$ is at the heart of our considerations. $\mathcal{I}_{\mu}^{\ \nu}$ also arises in the large-parameter limit of the finite conformal transformation (\ref{eq1}) or (\ref{eq4}).
\beq \label{eq25}
x^{'\mu} \rightarrow \frac{c^{\mu}}{c^2}+\frac{1}{c^2} \mathcal{I}^{\mu}_{\ \nu}(c)X^{\nu}+ O(\frac{1}{c^3})
\eeq
Thus a large-c conformal transformation is achieved
with a translation (by $c^{\mu}/c^2$), a scale transformation (by $1/c^2$), an improper Lorentz transformation (by $\mathcal{I}^{\mu}_{\ \nu}(c)$) and the inversion $X^{\nu}$. Since (\ref{eq1}) is connected to the identity, and the improper Lorentz transformation is not, it follows that the inversion is also an improper transformation.

We wonder whether there is an intrinsic explanation for the decoupling of Lorentz tensor/spinor indices from conformal transformations. 

\section*{Acknowledgment}
This work is supported in part by the National Science Council of R.O.C. under Grant number:
NSC98-2917-I-564-122(S.-H. H.); by the U.S. Department of Energy (DoE) under the cooperative
research agreements -05ER41360 (R.J.) and -91ER40676 (S-Y. P.).
